# Phenomenology of neutrino mixing in vacuum and matter

A Upadhyay[1] and M Batra[1]
School of Physics and Material Science
Thapar University, Patiala-147004.

E-mail: mbatra310@gmail.com

**Abstract:** During last fifty years, several past and currently going neutrino experiments like Super KamioKande, Sudbury Neutrino Observatory have not only resolved the mystery of solar and atmospheric neutrinos but also proved the massive nature of neutrinos. In view of important role that oscillation plays in determining the properties of neutrinos, detailed derivation for neutrino oscillation have been studied. Specifically the case of solar and the atmospheric neutrinos in vacuum and matter have been discussed. The phenomenology of neutrino oscillations is well established by theoretical and experimental grounds.

## 1. Introduction

Neutrinos are one of the least massive elementary particles in the set of building blocks of nature, they have no charge and belong to the family of leptons having half integral spin. The standard model predicts neutrinos to be mass less and escape easily but the flavor oscillation phenomenon leads to non-zero mass. The two contradictory theories led to need of further development of neutrino physics. Moreover the discrepancy in the amount of solar neutrino flux motivated the researchers to go deep into the situation. Combined Standard solar models [1] were proposed to obtain data on solar neutrino flux and various experiments were performed to resolve the mystery. Now a day's solar, atmospheric, reactor and accelerator based neutrino experiments confirmed that this mysterious particle change their flavors. This modification of neutrino properties (being massive and mixing with other flavor) requires an extension of the Standard Model. The confidence in such a solution is based on a prior exclusion of astrophysical or nuclear physic explanations; only in the past decade has such an outcome become strongly credible. Atmospheric neutrino anomaly is related to electron to muon neutrino ratio which reduces to half of the expected value and possible explanation is change of flavor from $\nu_\mu \Leftrightarrow \nu_\tau$, $\nu_\mu \Leftrightarrow \nu_\tau$. There are several key experiments which have been confirmed since the mystery of missing neutrino came into existence. Different types of neutrino beams like solar, atmospheric, reactor and accelerators have contributed to the present history of neutrino masses and mixing. Neutrinos are coming from various natural resources like interaction of cosmic rays, natural radioactivity, in the burning of stars but we here are more focused on the story of solar and atmospheric neutrinos. The disappearance of atmospheric and solar neutrino flux resulted in neutrino oscillation phenomenology. In this article, we mainly focus on various aspects of neutrino oscillation in vacuum and matter. Also the parameters which are affecting the neutrino oscillations are also included in our studies. Various experiments are analyzed in detail and their significance is mentioned.

Neutrino oscillations are periodic transitions between different flavor neutrinos in neutrino beams. In the quantum field theory the dependence of states on the time is given by Schrodinger equation.

$$i\frac{\partial |\Psi(t)>}{\partial t} = H|\Psi(t)> \tag{1.1}$$

Where H is the total Hamiltonian and the general solution of equation (1.1) is:

$$|\Psi(t)> = e^{-iHt}|\Psi(0)> \tag{1.2}$$

Where $|\Psi(0)>$ is the state at the initial time (t=0)

Let the initial state of flavor neutrino is $v_l$ ( l= e,μ,τ) and

$$|v_l> = \sum_{i=0}^{3} U_{li}^* |v_i> \tag{1.3}$$

Where $U_{li}^*$ is the unitary lepton mixing matrix and $|v_i>$ is the mass eigen state. The superposition of mass eigen states is called neutrino of flavor l. Thus we have at time (t=0)

$$|\Psi(0)> = |v_i> \tag{1.4}$$

As when we apply Hamiltonian on some wave function then we get the energy operator therefore taking into account that

$$H|v_i> = E_i|v_i>$$

Where $\quad E_i = \sqrt{p_i^2 + m_i^2}$

From eqn (1.2) & (1.3), we find the state of the left-handed neutrino at the time t ≥ 0, we have

$$|v_l>_t = e^{-iHt}|v_l> = \sum_{i=1}^{3} e^{-iE_i t} U_{li}^* |v_i> \tag{1.5}$$

Similarly, for the state of the right-handed antineutrino

$$|\overline{v_l}>_t = e^{-iHt}|\overline{v_l}> = \sum_{i=1}^{3} e^{-iE_i t} U_{li} |\overline{v_i}> \tag{1.6}$$

Generally, neutrino energies $E_i$ (i=1, 2, 3) are different. As neutrinos are having flavor and they oscillate into one another therefore we can write the amplitude of the transition $v_l \to v_{l'}$ during the time t

$$A(v_l \to v_{l'}) = \sum_{i=1}^{3} U_{il'} e^{-iE_i t} U_{li}^* \tag{1.7}$$

Where $U_{il'}$ is the amplitude of transition from the state $|v_i>$ into $|v_{l'}>$. $e^{-iE_i t}$ is the propagation in the state with definite mass. $U_{li}^*$ is the amplitude of transition from initial flavor state $|v_l>$ into the state of neutrino with definite mass $|v_i>$. Analogously, for the amplitude of the transition $\overline{v_l} \to \overline{v_{l'}}$ during the time t is given by:

$$A(\overline{v_l} \to \overline{v_{l'}}) = \sum_{i=1}^{3} U_{il'}^* e^{-iE_i t} U_{li}$$

Probability is the square of the amplitude, thus we can write it as:

$$P(v_l \to v_{l'}) = \left|\sum_{i=1}^{3} U_{il'} e^{-iE_i t} U_{li}^*\right|^2 \tag{1.8}$$

and

$$P(\overline{v_l} \to \overline{v_{l'}}) = \left|\sum_{i=1}^{3} U_{il'}^* e^{-iE_i t} U_{li}\right|^2 \tag{1.9}$$

From eqn (1.8) & (1.9) we can find possible relations between probabilities:

$$\sum_l P(v_l \to v_{l'}) = 1, \quad \sum_{l'} P(\overline{v_l} \to \overline{v_{l'}}) = 1$$
$$\sum_{l'} P(v_l \to v_{l'}) = 1, \quad \sum_l P(\overline{v_l} \to \overline{v_{l'}}) = 1$$

In quantum field theory, states of particles are characterized by their momentum, helicity, mass etc. Let us assume that a mixed neutrino state is characterized by their momentum p with $p_i = p$ and masses $m_i$

Neutrinos are having less mass but high energy

$$\frac{m_i^2}{p^2} \ll 1$$

Therefore we have,
$$E_i \cong p + \frac{m_i^2}{2p}$$

$$E_j \cong p + \frac{m_j^2}{2p}$$

Thus,
$$E_i - E_j = \frac{\Delta m_{ji}^2}{2p} \qquad (1.10)$$

Where $\Delta m_{ji}^2 = m_i^2 - m_j^2$

As $E \approx p$, therefore
$$E_i - E_j = \frac{\Delta m_{ji}^2}{2E} \qquad (1.11)$$

Let us suppose t is the difference of production and detection time for the ultra relativistic neutrinos and L is the distance between source of neutrino and the detector.

$$t \cong L$$

$$(E_i - E_j)t = \frac{\Delta m_{ji}^2}{2E} L$$

$$P(\nu_l \to \nu_{l'}) = \left| \sum_{i=1}^{3} U_{l'i} e^{-i\frac{\Delta m_{ji}^2 L}{2E}} U_{li}^* \right|^2 \qquad (1.12)$$

The unitary condition

$$\sum_i U_{l'i} U_{li}^* = \delta_{ll'}$$ gives convenient expression for the neutrino probability

$$P(\nu_l \to \nu_{l'}) = \left| \delta_{ll'} + \sum_{i \neq j} U_{il'} (e^{-i\frac{\Delta m_{ji}^2 L}{2E}} - 1) U_{li}^* \right|^2 \qquad (1.13)$$

Analogously, for the case of antineutrino we can write it as:

$$P(\bar{\nu}_l \to \bar{\nu}_{l'}) = \left| \delta_{ll'} + \sum_{i \neq j} U_{il'}^* (e^{-i\frac{\Delta m_{ji}^2 L}{2E}} - 1) U_{li} \right|^2 \qquad (1.14)$$

The transition probability $\nu_l \to \nu_{l'}$ can also be presented as:

$$P(\nu_l \to \nu_{l'}) = \sum_{i,k} U_{l'i} U_{l'k}^* U_{li}^* U_{lk} e^{-i\frac{\Delta m_{ji}^2}{2E}L}$$

$$= \sum_i |U_{l'i}|^2 |U_{li}|^2 + 2\text{Re} \sum_{i>k}(U_{l'i} U_{l'k}^* U_{li}^* U_{lk} e^{-i\frac{\Delta m_{ji}^2}{2E}L}) \qquad (1.15)$$

Further, from the unitary relation we can easily obtain the following relation

$$\sum_i |U_{l'i}|^2 |U_{li}|^2 = \delta_{ll'} - 2\,\text{Re}\sum_{i>k}(U_{l'i} U_{l'k}^* U_{li}^* U_{lk}) \qquad (1.16)$$

$$= \delta_{ll'} - 2\,\text{Re}\sum_{i>k}(U_{l'i} U_{l'k}^* U_{li}^* U_{lk}) + 2\,\text{Re} \sum_{i>k}(U_{l'i} U_{l'k}^* U_{li}^* U_{lk} e^{-i\frac{\Delta m_{ji}^2}{2E}L}$$

$$= \delta_{ll'} - 2\,\text{Re}\sum_{i>k}(U_{l'i} U_{l'k}^* U_{li}^* U_{lk} (1 - e^{-i\frac{\Delta m_{ki}^2 L}{2E}})$$

Finally for any complex a and b, Re(ab) = Re(a)Re(b) − Im(a)Im(b)

$$= \delta_{ll'} - 2\,\text{Re}\sum_{i>k}(U_{l'i}\,U^*_{l'k}\,U^*_{li}U_{lk})\,(1-\cos\frac{\Delta m^2_{ki}L}{2E} + i\,\text{Sin}\frac{\Delta m^2_{ki}L}{2E})$$

$$P(\nu_l \to \nu_{l'}) = \delta_{ll'} - 2\,\text{Re}\sum_{i>k}(U_{l'i}U^*_{l'i}\,U^*_{li}U_{lk})(1 - \cos\frac{\Delta m^2_{ki}L}{2E}) + 2\sum_{i>k}\text{Im}(U_{l'i}U^*_{l'k}\,U^*_{li}U_{lk})\sin\frac{\Delta m^2_{ki}L}{2E}) \quad (1.17)$$

Similarly, for the antineutrino oscillation probability becomes

$$P(\nu_{\bar{l}} \to \nu_{\bar{l'}}) = \delta_{ll'} - 2\,\text{Re}\sum_{i>k}(U_{l'i}U^*_{l'i}\,U^*_{li}U_{lk})(1 - \cos\frac{\Delta m^2_{ki}L}{2E}) - 2\sum_{i>k}\text{Im}(U_{l'i}U^*_{l'k}\,U^*_{li}U_{lk})\sin\frac{\Delta m^2_{ki}L}{2E}) \quad (1.18)$$

## 2. Two flavor oscillation probability in vacuum:

Let $\nu_e, \nu_\mu$ be the flavor eigen states and $\nu_1$, $\nu_2$ be the mass eigen states with masses $m_1$ and $m_2$, respectively and both are having momentum p. States in the flavor and mass bases are related by a mixing matrix U where U relates weak eigen state

$$\begin{pmatrix}\nu_e\\ \nu_\mu\end{pmatrix} = U\begin{pmatrix}\nu_1\\ \nu_2\end{pmatrix}$$

where

$$U = \begin{pmatrix}\cos\theta & \sin\theta\\ -\sin\theta & \cos\theta\end{pmatrix}$$

where θ is the mixing angle,

$$\begin{pmatrix}\nu_e\\ \nu_\mu\end{pmatrix} = \begin{pmatrix}\cos\theta & \sin\theta\\ -\sin\theta & \cos\theta\end{pmatrix}\begin{pmatrix}\nu_1\\ \nu_2\end{pmatrix}$$

$$|\nu_e(t=0)> = |\nu_e> = \cos\theta|\nu_1> + \sin\theta|\nu_2>$$
$$|\nu_\mu(t=0)> = |\nu_\mu> = -\sin\theta|\nu_1> + \cos\theta|\nu_2>$$

The weak eigen states are rotated by an angle θ with respect to mass eigen states $|\nu_1\rangle$ and $|\nu_2\rangle$ to allow mixing between $\nu_\mu$ and $\nu_e$. After some time t

$$|\nu_\mu(t=t)> = |\nu_\mu(t)> = -\sin\theta|\nu_1>e^{\frac{-iE_1 t}{\hbar}} + \cos\theta|\nu_2>e^{\frac{-iE_2 t}{\hbar}} = -\sin\theta|\nu_1>e^{-i(p+\frac{m_1^2}{2p})t/\hbar} + \cos\theta|\nu_2>e^{-i(p+\frac{m_2^2}{2p})t/\hbar}$$

Here we have used $E_1 = (p^2 + m_1^2)^{1/2}$ and $E_2 = (p^2 + m_2^2)^{1/2}$

$$|\nu_\mu(t)> = e^{-i\left(p+\frac{1}{2}\frac{m_1^2}{p}\right)t}(-\sin\theta|\nu_1> + \cos\theta|\nu_2>e^{-i(p+\frac{1}{2}\frac{m_1^2-m_2^2}{p})t})$$

To calculate the probability for a "pure" $\nu_e$ state to oscillate into a $\nu_\mu$ state, we must square the quantum mechanical amplitude that describes this transition.

$$P(\nu_e \to \nu_\mu) = |<\nu_\mu|\nu_e(t)>|^2$$

Where $<\nu_\mu| = \cos\theta<\nu_1| + \sin\theta<\nu_2|$

$$P(\nu_e \to \nu_\mu) = |<\nu_\mu|\nu_e(t)>|^2$$

$$= (e^{-iz}(-\sin\theta\cos\theta + \sin\theta\cos\theta\, e^{\frac{i\Delta m^2}{2p}x}))^2$$

$$= e^{iz-iz}\sin^2\theta\cos^2\theta\,(1 - e^{\frac{i\Delta m^2}{2p}x})(1 - e^{\frac{-i\Delta m^2}{2p}x})$$

Since the neutrino is relativistic, we can also make the substitution: $p = E_\nu$ and likewise we will make the substitution $x = L$.

$$P(\nu_e \rightarrow \nu_\mu) = \sin^2\theta \cos^2\theta\, (1-e^{\frac{i\Delta m^2}{2\,E_\nu}L})(1-e^{\frac{-i\Delta m^2}{2\,E_\nu}L})$$

$$P_{(\nu_e \rightarrow \nu_\mu)}(L,E) = \sin^2 2\theta \, \sin^2(1.27\, \Delta m^2\, \frac{L}{E_\nu}) \qquad (1.19)$$

where $\theta$ is the mixing angle which represents the amount of mixing between two mass eigen states, L is length of source from the detector, E is energy of neutrinos produced from the source. The mass squared difference values and mixing angle exists in nature. Physicists just probe the different mass eigen values and predict the mixing at which it occurs. Mixing angle dependence of the transition probability is expressed by $(\sin 2\theta)^2$. If we change from $\theta$ to $\pi/2-\theta$, the mixing angle dependence remains as such which confirms with degeneracy of oscillation probability for $\theta$ and $\pi/2-\theta$. Two possibilities here correspond to two physically different mixings for two mass eigen states if $\theta<\pi/4$, the electron neutrino is composed more of $\nu_1$ and if $\theta>\pi/4$ then muon neutrino is composed more of $\nu_2$. Moreover, transition to different flavor is not possible if $\Delta m^2\, L/2E\ll 1$ which led us to goto survival probability Direct information about the mixing angle can be obtained from average neutrino oscillation probability $\langle P(\nu_\alpha \rightarrow \nu_\beta)\rangle = \frac{1}{2}(\sin 2\theta)^2$.

### 3. Three Flavor Oscillation Probability in Vacuum:

In case of three flavor neutrino oscillation, standard parameterization of mixing matrix can be achieved by using three vectors and performing Euler rotations introduces three mixing angles and one complex phase factor. For three flavor and three mass eigen states it can be written as:

$$\begin{pmatrix}\nu_1\\\nu_2\\\nu_3\end{pmatrix} = U \begin{pmatrix}\nu_e\\\nu_\mu\\\nu_\tau\end{pmatrix}$$

Where U is unitary mixing matrix. Consider the unitary 3×3 mixing matrix for Dirac neutrinos and introduce the standard parameters (three mixing angles and one phase) which characterizes it. Three normalized vectors are defined as

$$|i> \qquad (i = 1, 2, 3)$$
$$\langle i|k\rangle = \delta_{ik}$$

The first Euler rotation performed at the angle $\theta_{12}$ around the vector $|3>$ produces new orthogonal and normalized vectors as:

$$|1>^{(1)} = c_{12}|1> + s_{12}|2>$$
$$|2>^{(1)} = -s_{12}|1> + c_{12}|2>$$
$$|3>^{(1)} = |3>$$

Where $c_{12} = \cos\theta_{12}$ and $s_{12} = \sin\theta_{12}$,

$$|\nu>^{(1)} = U^{(1)}|\nu> \quad \text{such that}$$

$$|\nu>^{(1)} = \begin{pmatrix}|1^{(1)}>\\|2^{(1)}>\\|3^{(1)}>\end{pmatrix} \qquad \text{and} \qquad |\nu> = \begin{pmatrix}|1>\\|2>\\|3>\end{pmatrix}$$

$$U^{(1)} = \begin{pmatrix} c_{12} & s_{12} & 0 \\ -s_{12} & c_{12} & 0 \\ 0 & 0 & 1 \end{pmatrix}$$

Second rotation at the angle $\theta_{13}$ around vector the $|2>^{(1)}$ introduces the CP phase $\delta$,

$$|1>^{(2)} = c_{13}|1>^{(1)} | + s_{13}e^{-i\delta}|3>^{(1)}$$

$$|2>^{(2)} = |2>^{(1)}$$

$$|3>^{(2)} = -s_{13}e^{-i\delta}|1>^{(1)} + c_{13}|3>^{(1)}$$

In the matrix $|v>^{(2)} = U^{(2)}|v>^{(1)}$ where $U^{(2)} = \begin{pmatrix} c_{13} & 0 & s_{13}e^{-i\delta} \\ 0 & 1 & 0 \\ -s_{13}e^{-i\delta} & 0 & c_{13} \end{pmatrix}$

Similarly rotation around vector $|1>^{(2)}$ at the angle $\theta_2$

$$|1>^{mix} = |1>^{(2)}$$

$$|2>^{mix} = c_{23}|2>^{(2)} + s_{23}|3>^{(2)}$$

$$|3>^{mix} = -s_{23}|2>^{(2)} + c_{23}|3>^{(2)}$$

$$|v^{mix}> = U^{(3)}|v>^{(2)}$$

$$U(3) = \begin{pmatrix} 1 & 0 & 0 \\ 0 & c_{23} & s_{23} \\ 0 & -s_{23} & c_{23} \end{pmatrix}$$

$$|v^{mix}> = U|v> \text{ Where } U = U^{(3)}U^{(2)}U^{(1)}$$

$$U = \begin{pmatrix} 1 & 0 & 0 \\ 0 & c_{23} & s_{23} \\ 0 & -s_{23} & c_{23} \end{pmatrix} \begin{pmatrix} c_{13} & 0 & s_{13}e^{-i\delta} \\ 0 & 1 & 0 \\ s_{13}e^{-i\delta} & 0 & c_{13} \end{pmatrix} \begin{pmatrix} c_{12} & s_{12} & 0 \\ -s_{12} & c_{12} & 0 \\ 0 & 0 & 1 \end{pmatrix}$$

$$= \begin{pmatrix} c_{13} & 0 & s_{13}e^{-i\delta} \\ -s_{13}s_{23}e^{-i\delta} & c_{23} & s_{23}c_{13} \\ -c_{23}s_{13}e^{-i\delta} & -s_{23} & c_{23}c_{13} \end{pmatrix} \begin{pmatrix} c_{12} & s_{12} & 0 \\ -s_{12} & c_{12} & 0 \\ 0 & 0 & 1 \end{pmatrix}$$

$$U = \begin{pmatrix} c_{13}c_{12} & c_{13}s_{12} & s_{13}e^{-i\delta} \\ -c_{23}s_{12} - c_{12}s_{13}s_{23}e^{i\delta} & c_{12}c_{23} - s_{12}s_{13}s_{23}e^{i\delta} & c_{13}s_{23} \\ s_{23}s_{12} - s_{13}c_{12}c_{23}e^{i\delta} & c_{12}s_{23} - s_{13}c_{23}s_{12}e^{i\delta} & c_{13}c_{23} \end{pmatrix}$$

The phase $\delta$ is responsible for effects of the CP violation which can take values from 0 to $2\pi$. The mixing angles are parameters which can take values in the ranges $0 \leq \theta_{12} \leq \pi$, $0 \leq \theta_{13} \leq \pi$, $0 \leq \theta_{23} \leq \pi$. For three neutrino oscillation in vacuum, all real parts of the quadratic products of elements of the mixing matrix entering in the three-neutrino oscillation probabilities is given as: $Re(U_{l'i} U_{l'k}^* U_{li}^* U_{lk})$. The individual probability expression for three neutrino flavors changing into others can be achieved by solving for individual matrix elements. For electron to muon transition: $P(v_e \to v_\mu)$ can be calculated as:

$$Re(U_{l'i} U_{l'k}^* U_{li}^* U_{lk}) = (U_{22}U_{21}^*U_{12}^*U_{11}) \text{ for } l' = 2, l = 1, i = 2, k = 1$$

$$= (c_{23}c_{12} - s_{13}s_{23}s_{12}e^{i\delta})(-c_{23}s_{12} - s_{13}s_{23}c_{12}e^{-i\delta})(c_{13}s_{12})(c_{13}c_{12})$$

$$= c_{13}^2c_{12}s_{12}(-c_{23}^2c_{12}s_{12} - c_{23}c_{12}s_{13}s_{23}c_{12}e^{-i\delta} + c_{23}s_{23}s_{13}s_{12}^2e^{-i\delta} + s_{12}c_{12}s_{13}^2s_{23}^2)$$

$$= -\frac{1}{4}c_{13}^2\sin2\theta_{12}(c_{23}^2\sin2\theta_{12}+s_{13}c_{12}^2\sin2\theta_{23}e^{-i\delta} - s_{13}s_{12}^2\sin2\theta_{23}e^{-i\delta} - s_{23}^2s_{13}^2\sin2\theta_{12})$$

$$= -\frac{1}{4}c_{13}^2\sin2\theta_{12}(c_{23}^2\sin2\theta_{12}+s_{13}c_{12}^2\sin2\theta_{23}\cos\delta_{13} - s_{13}s_{12}^2\sin2\theta_{23}\cos\delta_{13} - s_{23}^2s_{13}^2\sin2\theta_{12})$$

$$= -\frac{1}{4}c_{13}^2\sin2\theta_{12}[\sin2\theta_{12}(c_{23}^2 - s_{23}^2s_{13}^2)+ \sin2\theta_{23}s_{13} \cos\delta_{13}(\cos^2\theta_{12} - \sin^2\theta_{12})]$$

$$= -\frac{1}{4}c_{13}^2\sin2\theta_{12}[\sin2\theta_{12}(c_{23}^2 - s_{23}^2s_{13}^2)+\cos2\theta_{12} \sin2\theta_{23}s_{13} \cos\delta_{13}]$$

$\text{Re}(U_{l'i} U_{l'k}^* U_{li}^* U_{lk}) = (U_{23}U_{22}^*U_{13}^*U_{12})$

$$= (c_{13}s_{23})(c_{23}c_{12} - s_{13}s_{23}s_{12}e^{-i\delta})(s_{13}e^{i\delta})(c_{13}s_{12})$$

$$= c_{13}^2s_{13}s_{12}s_{23}e^{i\delta}(c_{23}c_{12} - s_{13}s_{23}s_{12}e^{-i\delta})$$

$$= c_{13}^2s_{13}s_{12}s_{23}(c_{23}c_{12}e^{i\delta} - s_{13}s_{23}s_{12})$$

$$= -c_{13}^2s_{13}s_{12}s_{23}(s_{13}s_{23}s_{12} - c_{23}c_{12}\cos\delta_{13})$$

$\text{Re}(U_{l'i} U_{l'k}^* U_{li}^* U_{lk})= (U_{23}U_{21}^*U_{13}^*U_{11})$

$$= (c_{13}s_{23})(-c_{23}s_{12} - s_{13}s_{23}c_{12}e^{-i\delta})(s_{13}e^{i\delta})(c_{13}c_{12})$$

$$= c_{13}^2s_{13}c_{12}s_{23}e^{i\delta}(-c_{23}s_{12} - s_{13}s_{23}c_{12}e^{-i\delta})$$

$$= c_{13}^2s_{13}c_{12}s_{23}(-c_{23}s_{12}e^{i\delta} - s_{13}s_{23}c_{12})$$

$$= -c_{13}^2s_{13}c_{12}s_{23}(c_{23}s_{12}e^{i\delta} + s_{13}s_{23}c_{12})$$

Similar expressions can be derived in table 1.1.

| Probability | i, k | $\text{Re}(U_{l'i} U_{l'k}^* U_{li}^* U_{lk})$ |
|---|---|---|
| $P(\nu_e \to \nu_\mu)$ | i=2, k=1 | $-\frac{1}{4}c_{13}^2\sin 2\theta_{12}[\sin 2\theta_{12}(c_{23}^2-s_{23}^2 s_{13}^2)+\cos 2\theta_{12}\sin 2\theta_{23}s_{13}\cos\delta_{13}]$ |
| | i=3, k=2 | $c_{13}^2 s_{13} s_{12} s_{23}(s_{13}s_{23}s_{12}-c_{23}c_{12}\cos\delta_{13})$ |
| | i=3, k=1 | $c_{13}^2 s_{13} c_{12} s_{23}(c_{23}s_{12}e^{i\delta}+s_{13}s_{23}c_{12})$ |
| $P(\nu_e \to \nu_\tau)$ | i=2, k=1 | $\frac{1}{4}c_{13}^2\sin 2\theta_{12}[\sin 2\theta_{12}(c_{23}^2 s_{13}^2-s_{23}^2)+\cos 2\theta_{12}\sin 2\theta_{23}s_{13}\cos\delta_{13}]$ |
| | i=3, k=2 | $-c_{13}^2 s_{13} s_{12} c_{23}(s_{13}c_{23}s_{12}+s_{23}c_{12}\cos\delta_{13})$ |
| | i=3, k=1 | $-c_{13}^2 s_{13} c_{12} c_{23}(s_{13}c_{23}c_{12}-s_{23}s_{12}\cos\delta_{13})$ |
| $P(\nu_e \to \nu_e)$ | i=2, k=1 | $\frac{1}{4}c_{13}^4\sin^2 2\theta_{12}$ |
| | i=3, k=2 | $\frac{1}{4}s_{12}^2\sin^2 2\theta_{13}$ |
| | i=3, k=1 | $\frac{1}{4}c_{12}^2\sin^2 2\theta_{13}$ |
| $P(\nu_\mu \to \nu_\tau)$ | i=2, k=1 | $\frac{1}{16}\sin^2 2\theta_{12}\sin^2 2\theta_{23}(1+s_{13}^2)^2-\frac{1}{4}(\sin^2 2\theta_{12}+\sin^2 2\theta_{23})s_{13}^2-\frac{1}{16}\sin 4\theta_{12}\sin 4\theta_{23}(1+s_{13}^2)s_{13}\cos\delta_{13}+\frac{1}{4}\sin^2 2\theta_{12}\sin^2 2\theta_{23}s_{13}^2\cos^2\delta_{13}$ |
| | i=3, k=2 | $\frac{-1}{4}\sin 2\theta_{23}c_{13}^2[\sin 2\theta_{23}(c_{12}^2-s_{12}^2 s_{13}^2)+\sin 2\theta_{12}\cos 2\theta_{23}s_{13}\cos\delta_{13}]$ |
| | i=3, k=1 | $\frac{1}{4}\sin 2\theta_{23}c_{13}^2 [\sin 2\theta_{23}(c_{12}^2 s_{13}^2-s_{12}^2)+\sin 2\theta_{12}\cos 2\theta_{23}s_{13}\cos\delta_{13}]$ |
| $P(\nu_\mu \to \nu_\mu)$ | i=2, k=1 | $\frac{1}{4}\sin^2 2\theta_{12}(c_{23}^4+s_{23}^4 s_{13}^2)+\frac{1}{4}(1-\frac{1}{2}\sin^2 2\theta_{12})\sin^2 2\theta_{23}s_{13}^2+\frac{1}{4}\sin 4\theta_{12}\sin 2\theta_{23}(c_{23}^2-s_{23}^2 s_{13}^2)s_{13}\cos\delta_{13}-\frac{1}{4}\sin^2 2\theta_{12}\sin^2 2\theta_{23}s_{13}^2\cos^2\delta_{13}$ |
| | i=3, k=2 | $s_{23}^2 c_{13}^2(c_{12}^2 c_{23}^2+s_{12}^2 s_{23}^2 s_{13}^2-\frac{1}{2}\sin 2\theta_{12}\sin 2\theta_{23}s_{13}\cos\delta_{13})$ |
| | i=3, k=1 | $s_{23}^2 c_{13}^2(s_{12}^2 c_{23}^2+c_{12}^2 s_{23}^2 s_{13}^2+\frac{1}{2}\sin 2\theta_{12}\sin 2\theta_{23}s_{13}\cos\delta_{13})$ |
| $P(\nu_\tau \to \nu_\tau)$ | i=2, k=1 | $\frac{1}{4}\sin^2 2\theta_{12}(s_{23}^4+c_{23}^4 s_{13}^2)+\frac{1}{4}(1-\frac{1}{2}\sin^2 2\theta_{12})\sin^2 2\theta_{23}s_{13}^2+\frac{1}{4}\sin 4\theta_{12}\sin 2\theta_{23}(s_{23}^2-c_{23}^2 s_{13}^2)s_{13}\cos\delta_{13}-\frac{1}{4}\sin^2 2\theta_{12}\sin^2 2\theta_{23}s_{13}^2\cos^2\delta_{13}$ |
| | i=3, k=2 | $c_{23}^2 c_{13}^2\ (c_{12}^2 s_{23}^2+s_{12}^2 c_{23}^2 s_{13}^2+\frac{1}{2}\sin 2\theta_{12}\sin 2\theta_{23}s_{13}\cos\delta_{13})$ |
| | i=3, k=1 | $c_{23}^2 c_{13}^2(s_{12}^2 s_{23}^2+c_{12}^2 c_{23}^2 s_{13}^2-\frac{1}{2}\sin 2\theta_{12}\sin 2\theta_{23}s_{13}\cos\delta_{13})$ |

From the table it is clear that there are three types of mass mixing, $\Delta m_{21}^2$, $\Delta m_{31}^2$, $\Delta m_{32}^2$ where only two are independent.

$$\Delta m_{32}^2 + \Delta m_{21}^2 - \Delta m_{31}^2 = 0$$

Thus neutrino oscillations are only sensitive to mass squared difference in spite of actual mass. The solar experiments have inferred the sign of $\Delta m^2$ from MSW effect but sign of mass squared difference in atmospheric is not known and the condition observed at experiments, $\Delta m_{sol}^2 \ll \Delta m_{atmos}^2$, predicts that the two types of neutrino mixing can occur. One is normal hierarchy having two light states and one heavier ($m_1 \ll m_2 < m_3$) but for inverted hierarchy, $m_3$ is lightest state which assume masses in order ($m_3 \ll m_1 < m_2$). The two hierarchies are shown in figure below.[2]

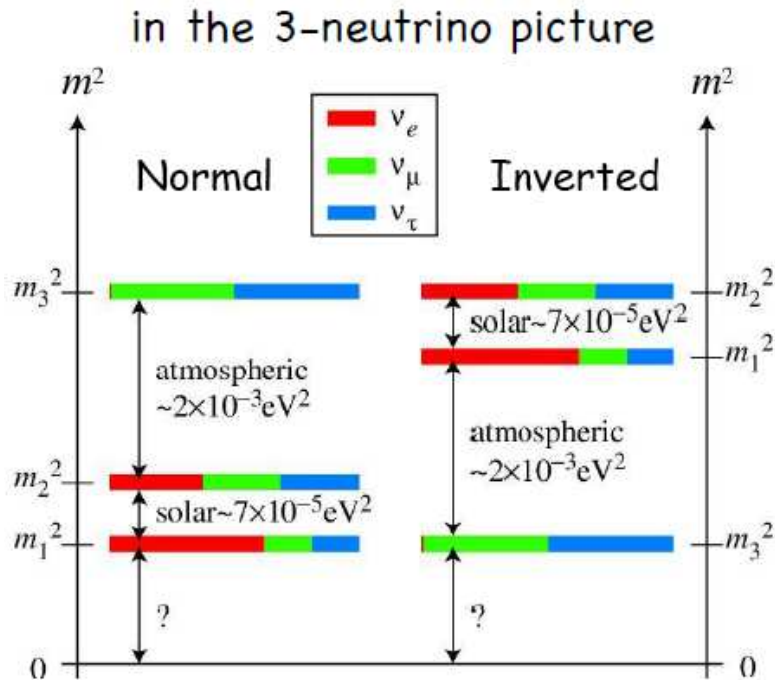

Figure 3.1 -Normal and inverted hierarchy

*3.1 Atmospheric Oscillation Probability:*

Atmospheric neutrinos are created by interactions of primary cosmic rays with nuclei in atmosphere. The neutrinos generate upward-going and horizontal muons through decays. The phenomena of atmospheric neutrino oscillations arise from the deficit in upward and downward going muon neutrino. Let us calculate probability for electron neutrino changing into muon neutrino, as we already discussed that there are three possible cases for electron neutrino changing into muon neutrino i.e. from one mass eigen state to another.

$P(\nu_e \to \nu_\mu) = \delta_{12} - 4[-\frac{1}{4}c_{13}^2\sin2\theta_{12}[\sin2\theta_{12}(c_{23}^2 - s_{23}^2 s_{13}^2) + \cos2\theta_{12}\sin2\theta_{23}s_{13}\cos\delta_{13}]$

$\sin\frac{\Delta m_{21}^2 L}{2E} + \delta_{32} - 4[-c_{13}^2 s_{13} s_{12} s_{23}(s_{13}s_{23}s_{12} - c_{23}c_{12}\cos\delta_{13})]\sin\frac{\Delta m_{32}^2 L}{4E} + \delta_{31} - 4[-c_{13}^2 s_{13} c_{12} s_{23}(c_{23}s_{12}e^{i\delta} +$

$s_{13}s_{23}c_{12})]\sin\frac{\Delta m_{31}^2 L}{4E}$

For atmospheric neutrinos,

$$\Delta m_{32}^2 \cong \Delta m_{31}^2 \cong \Delta m_{atm}^2$$

$$\Delta m_{21}^2 \cong \Delta m_{sol}^2 \cong 0$$

$P(\nu_e \to \nu_\mu) = -4(-s_{12}^2 s_{23}^2 s_{13}^2 c_{13}^2 + s_{12}s_{23}s_{13}c_{23}c_{12}c_{13}^2\cos\delta_{13} - s_{13}^2 c_{13}^2 - s_{12}s_{23}s_{13}c_{23}c_{12}c_{13}^2\cos\delta_{13})\sin\frac{\Delta m_{atm}^2 L}{4E}$

$\qquad = 4[s_{23}^2 s_{13}^2 c_{13}^2 (s_{12}^2 + c_{12}^2)]\sin\frac{\Delta m_{atm}^2 L}{4E}$

$\qquad = 4\sin^2\theta_{23}\cos^2\theta_{13}\sin^2\theta_{13}\sin\frac{\Delta m_{atm}^2 L}{4E}$

$P(\nu_e \to \nu_\mu) = \sin^2(2\theta_{13})\sin^2(\theta_{23})\sin\frac{\Delta m_{atm}^2 L}{4E}$ \hfill (1.19)

$P(\nu_e \to \nu_\tau) = \sin^2(2\theta_{13})\cos^2(\theta_{23})\sin\frac{\Delta m_{atm}^2 L}{4E}$ \hfill (1.20)

The survival probability of electron neutrinos is

$P(\nu_e \to \nu_e) = 1 - [P(\nu_e \to \nu_\mu) + P(\nu_e \to \nu_\tau)]$

$\qquad = 1 - [\sin^2(2\theta_{13})\sin^2(\theta_{23})\sin\frac{\Delta m_{atm}^2 L}{4E} + \sin^2(2\theta_{13})\cos^2(\theta_{23})\sin\frac{\Delta m_{atm}^2 L}{4E}]$

$\qquad = 1 - [\sin^2(2\theta_{13})\sin\frac{\Delta m_{atm}^2 L}{4E}\{\sin^2(\theta_{23}) + \cos^2(\theta_{23})\}]$

$P(\nu_e \to \nu_e) = 1 - \sin^2(2\theta_{13})\sin\frac{\Delta m_{atm}^2 L}{4E}$ \hfill (1.21)

In the case of oscillation in vacuum, electron neutrinos going into muon neutrinos and vice-versa probability remains the same.

Therefore in case of vacuum

$$P(\nu_e \to \nu_\mu) = P(\nu_\mu \to \nu_e)$$

Thus we can write it as:

$$P(\nu_\mu \to \nu_e) = \sin^2(2\theta_{13})\sin^2(\theta_{23})\sin\frac{\Delta m_{atm}^2 L}{4E} \qquad (1.22)$$

Similarly we can calculate the probability for muon neutrinos changing into tau neutrinos

$$P(\nu_\mu \to \nu_\tau) = \sin^2(2\theta_{23})\cos^4(\theta_{13})\sin\frac{\Delta m_{atm}^2 L}{4E} \qquad (1.23)$$

Now we will calculate the survival probability of muon neutrinos i.e.

$P(\nu_\mu \to \nu_\mu) = 1 - [P(\nu_\mu \to \nu_e) + P(\nu_\mu \to \nu_\tau)]$

$\qquad = 1 - [\sin^2(2\theta_{13})\sin^2(\theta_{23})\sin\frac{\Delta m_{atm}^2 L}{4E} + \sin^2(2\theta_{23})\cos^4(\theta_{13})\sin\frac{\Delta m_{atm}^2 L}{4E}]$

$\qquad = 1 - \sin^2(2\theta_{13})\sin^2(\theta_{23}) + \sin^2(2\theta_{23})\cos^4(\theta_{13})\sin\frac{\Delta m_{atm}^2 L}{4E}$ \hfill (1.24)

These are the six probability terms from where we can find the probability for one neutrinos changing into another depending upon their flavor.

*3.2 Solar Neutrino Probability*

The difference in the number of solar neutrinos predicted from solar models and number of neutrinos flowing through earth led to solar neutrino problem this created the solar neutrinos as the target for researchers as it can provide more elaborated picture of stellar evolution and energy resources. The chlorine, Kamiokande, Super-Kamiokande, GALLEX, SAGE, GNO, and SNO experiments measured solar neutrino events to find solution of solar neutrino problem. The various experiments are focused to measure solar neutrino flux. Solar neutrino events can also be analyzed by a Monte-Carlo simulation study of uncertainties which made use of fluxes from 100 standard solar models. The following figure [3] shows solar energy spectrum of neutrinos coming from various sources like pp chain and CNO cycle.

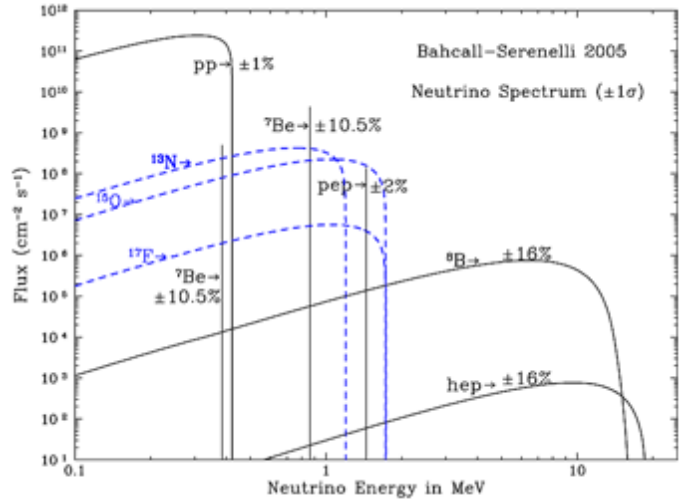

**Figure 3.2 Neutrino Energy Spectrum**

The fact that neutrino oscillations being observed in atmospheric neutrino, further strengthens the case that oscillations occur for solar neutrinos too. Solar neutrino experimental data constrains that mass squared difference $\Delta m_{21}^2$ is only taken where other mass differences are neglected.

$P(\nu_e \to \nu_\mu) = \delta_{12} - 4[-\frac{1}{4}c_{13}^2\sin 2\theta_{12}[\sin 2\theta_{12}(c_{23}^2 - s_{23}^2 s_{13}^2) + \cos 2\theta_{12} \sin 2\theta_{23} s_{13} \cos\delta_{13}] \sin\frac{\Delta m_{21}^2 L}{2E}$

$P(\nu_e \to \nu_\mu) = -4[-\frac{1}{4}c_{13}^2\sin 2\theta_{12}[\sin 2\theta_{12}(c_{23}^2 - s_{23}^2 s_{13}^2) + \cos 2\theta_{12} \sin 2\theta_{23} s_{13} \cos\delta_{13}] \sin\frac{\Delta m_{21}^2 L}{2E}$

$= [\sin^2 2\theta_{12}\cos^2\theta_{13}(\cos^2\theta_{23} - \sin^2\theta_{23}\sin^2\theta_{13}) + \frac{1}{4}\sin 4\theta_{12} \sin 2\theta_{13} \sin 2\theta_{23}\cos\theta_{13}\cos\delta_{13}]\sin\frac{\Delta m_{21}^2 L}{2E}$.

For oscillation in vacuum δ =0, therefore $\cos\delta_{13} = 1$

$P(\nu_e \to \nu_\mu) = [\sin^2 2\theta_{12}\cos^2(\theta_{13})(\cos^2\theta_{23} - \sin^2\theta_{23}\sin^2\theta_{13}) + \frac{1}{4}(\sin 4\theta_{12}\sin 2\theta_{13}\cos\theta_{13})\sin\frac{\Delta m_{21}^2 L}{2E}$.

Similarly,

$P(\nu_e \to \nu_\tau) = \delta_{12} - 4[\frac{1}{4}c_{13}^2\sin 2\theta_{12}\{\sin 2\theta_{12}(c_{23}^2 s_{13}^2 - s_{23}^2) + \cos 2\theta_{12} \sin 2\theta_{23} s_{13} \cos\delta_{13}\}] \sin\frac{\Delta m_{21}^2 L}{4E}$

$= \sin^2 2\theta_{12} \cos^2\theta_{13}(\cos^2\theta_{23}\sin^2\theta_{13} - \sin^2\theta_{23}) + \sin 2\theta_{12}\cos 2\theta_{12} \cos^2\theta_{13} \sin\theta_{13} \sin 2\theta_{23} \cos\delta_{13}] \sin\frac{\Delta m_{21}^2 L}{2E}$

$$= [-\sin^2 2\theta_{12}\cos^2\theta_{13}(\cos^2\theta_{23}\sin^2\theta_{13}- \sin^2\theta_{23}) -\frac{1}{4}\sin 4\theta_{12}\ \sin 2\theta_{13}\ \sin 2\theta_{23}\cos\theta_{13}\cos\delta_{13}]\sin\frac{\Delta m_{21}^2 L}{2E}$$

$$= [-\sin^2 2\theta_{12}\cos^2\theta_{13}(\cos^2\theta_{23}\sin^2\theta_{13}- \sin^2\theta_{23}) -\frac{1}{4}\sin 4\theta_{12}\ \sin 2\theta_{13}\ \sin 2\theta_{23}\cos\theta_{13}]\sin\frac{\Delta m_{21}^2 L}{2E}$$

Survival probability,

$$P(\nu_e \rightarrow \nu_e)=1-[\sin^2 2\theta_{12}\cos^2\theta_{13}(\cos^2\theta_{23}-\sin^2\theta_{23}\sin^2\theta_{13})+\frac{1}{4}\sin 4\theta_{12}$$

$$\sin 2\theta_{13}\sin 2\theta_{23}\cos\theta_{13}-\sin^2 2\theta_{12}\cos^2\theta_{13}(\cos^2\theta_{23}\sin^2\theta_{13}-\sin^2\theta_{23})-\frac{1}{4}\sin 4\theta_{12}\sin 2\theta_{13}\ \sin 2\theta_{23}\cos\theta_{13}\cos\delta_{13}]\ \sin\frac{\Delta m_{21}^2 L}{2E}$$

$$=1 - [\ \sin^2 2\theta_{12}\cos^2\theta_{13}(\cos^2\theta_{23}- \sin^2\theta_{23}\ \sin^2\theta_{13}) -\sin^2 2\theta_{12}\cos^2\theta_{13}(\cos^2\theta_{23}\sin^2\theta_{13}- \sin^2\theta_{23})]\ \sin\frac{\Delta m_{21}^2 L}{2E}$$

$$=1 - [\ \sin^2 2\theta_{12}\cos^2\theta_{13}(\cos^2\theta_{23}- \sin^2\theta_{23}\ \sin^2\theta_{13} - \cos^2\theta_{23}\sin^2\theta_{13}+ \sin^2\theta_{23})]\ \sin\frac{\Delta m_{21}^2 L}{2E}$$

$$=1 - [\ \sin^2 2\theta_{12}\cos^2\theta_{13}\{\ \cos^2\theta_{23}(1- \sin^2\theta_{13}) + \sin^2\theta_{23}(1- \sin^2\theta_{13})\}]\ \sin\frac{\Delta m_{21}^2 L}{2E}$$

$$=1 - [\ \sin^2 2\theta_{12}\cos^2\theta_{13}\{\ \cos^2\theta_{23}\ \cos^2\theta_{13} + \sin^2\theta_{23}\ \cos^2\theta_{13}\}]\ \sin\frac{\Delta m_{21}^2 L}{2E}$$

$$=1 - [\ \sin^2 2\theta_{12}\cos^2\theta_{13}\ \cos^2\theta_{13}(\cos^2\theta_{23} + \sin^2\theta_{23})]\ \sin\frac{\Delta m_{21}^2 L}{2E}$$

$$= 1 - [\ \sin^2 2\theta_{12}\cos^4\theta_{13}]\ \sin\frac{\Delta m_{21}^2 L}{2E}$$

## 4. Neutrino oscillation in matter:

Since neutrinos are weakly interacting, they might interact with matter either through charged-current (CC) and neutral current (NC) interactions [4]. For the charged current interactions only the electrons participate via $W^\pm$ exchange. Mikheev and Smirnov [5] noticed resonance behavior for specific oscillation and matter density parameters. Therefore the probabilities for neutrino oscillation differ from their vacuum counterparts. Neutral current flavor interactions can occur for any type of neutrino flavor. Moreover, neutral current interaction leads to addition of an extra term in Hamiltonian for flavor oscillation such a term produces a shift in eigen values but charged current interactions give the contribution not in the form of change of eigen states only adds to Hamiltonian an energy proportional to $V=\sqrt{2}\ G_F N_e$ where $G_F$ =Fermi Coupling Constant and $N_e$=Number of electrons per unit volume.

Neutrinos are produced in flavor eigen states, $|\upsilon_\alpha\rangle$, ($\alpha$=e,μ,τ) created by interaction of weak gauge bosons with the charged leptons between the source, production point of the neutrinos and the detector. The state $|\psi(t)\rangle$ of neutrinos with momentum p satisfies equation

$$i\frac{\partial}{\partial t} = H_0 |\psi(t)\rangle$$

When $H_0$ is free Hamiltonian. The state $|\psi(t)\rangle$ can be expanded over the total system of states of flavor neutrinos $\nu_l$ with momentum p,

$$|\psi(t)\rangle = \sum a_l(t)|\nu_l\rangle$$

Here
$$|\nu_l\rangle = \sum_i U_{li}^* |\nu_i\rangle$$

$H_0|v_i\rangle = E_i|v_i\rangle$, $H_0|v_i\rangle = E_i|v_i\rangle$, $E_i = \sqrt{p_i^2 + m_i^2} \simeq p + \dfrac{m_i^2}{2E}$ and $a_l(t) = \langle v_l|\psi(t)\rangle$ is the amplitude of probability to find $v_l$ in state which is described by $|\psi(t)\rangle$.

Therefore,

$$i\frac{\partial a_l^{'}(t)}{\partial t} = \sum_l \langle v_l^{'}|H_0|a_l(t)\rangle$$

Where,
$$\langle v_l^{'}|v_l\rangle = U_{l'l}$$

$$\langle v_l|v_{l'}\rangle = U_{l'l}^*$$

Taking into account this relation, for the free Hamiltonian in the flavor representation we have the following expression:

$$\langle v_{l'}|H_0|v_l\rangle = \sum_l U_{l'i} E_i U_{li}^* \sim p + \sum_l U_{l'i}\frac{m_i^2}{2E}U_{li}^*$$

Therefore neutrino evolution equation in the flavor representation:

$$i\frac{\partial a(t)}{\partial t} = U\frac{m^2}{2E}U^\dagger a(t)$$

Let us introduce the function
$$a^{'}(t) = U^\dagger a(t)$$

We find that the function $a^{'}(t)$ satisfies the following equation, Multiplying by $U^\dagger$ on both sides then

$$i\frac{\partial}{\partial t}a^{'}(t) = \frac{m^2}{2E}a'(t)$$

It is obvious that the solution of equation has the form: $a'(t) = e^{-i\frac{m^2}{2E}(t-t_0)}a'(t_0)$ where $a'(t_0)$ is the wave-function at the initial time $t_0$.

As the neutrinos propagate in matter, electron neutrino plays a very special role due to coherent forward scattering of neutrinos from electrons so that it leads to an additional contribution in oscillation probability. The forward scattering of electrons of matter with scattering of electrons of matter with neutrinos is called charge current interaction. Therefore the probabilities for neutrino oscillation differ from their vacuum counterparts. Charged current interaction can give contribution only to the process of elastic scattering of $v_e$ on electrons can give contribution to the process of elastic scattering of $v_e$ on electrons via $W^\pm$ exchange. Moreover, Neutral current interaction leads to addition of an extra term in Hamiltonian for flavor oscillation probability. For the low energy, an effective Hamiltonian of the neutrino interaction obtained from the diagonal matrix element $\langle pmat|H_I^{CC}|pmat\rangle$ where $H_I^{cc} = \dfrac{G_F}{\sqrt{2}}2\overline{v_{eL}}(x)\gamma_\alpha v_{eL}\overline{e}(x)\gamma^\alpha(1-\gamma_5)e(x)$ and the vector $|pmat\rangle = |p\rangle|mat\rangle$. Now substituting the Hamiltonian we obtain:

$$\left\langle pmat \left| \frac{G_F}{\sqrt{2}}2\overline{v_{eL}}(x)\gamma_\alpha v_{eL}\overline{e}(x)\gamma^\alpha(1-\gamma_5)e(x) \right| pmat \right\rangle$$

$$= \langle p|\langle mat|\frac{G_F}{\sqrt{2}} 2\overline{v_{eL}}(x)\gamma_\alpha v_{eL}\overline{e}(x)\gamma^\alpha(1-\gamma_5)e(x)|p\rangle|mat\rangle$$

$$\langle p|\langle mat|\frac{G_F}{\sqrt{2}} 2\overline{v_{eL}}(x)\gamma_\alpha v_{eL}\overline{e}(x)\gamma^\alpha - \gamma^\alpha\gamma_5)e(x)|p\rangle|mat\rangle =$$

$$\langle p|\langle mat|\frac{G_F}{\sqrt{2}} 2\overline{v_{eL}}(x)\gamma_\alpha v_{eL}\overline{e}(x)\gamma^\alpha|p\rangle|mat\rangle - \langle p|\langle mat|\frac{G_F}{\sqrt{2}} 2\overline{v_{eL}}(x)\gamma_\alpha v_{eL}\overline{e}(x)\gamma^\alpha\gamma_5 e(x)|p\rangle|mat\rangle$$

But $\langle p|\langle mat|\frac{G_F}{\sqrt{2}} 2\overline{v_{eL}}(x)\gamma_\alpha v_{eL}\overline{e}(x)\gamma^\alpha\gamma_5)e(x)|p\rangle|mat\rangle = 0$ as for unpolarized matter $\langle mat|\overline{e}(x)\gamma^\alpha\gamma_5)e(x)|mat\rangle = 0$

Also $\langle mat|\overline{e}(x)\gamma^\alpha e(x)|p\rangle|mat\rangle = \langle mat|\overline{e}(x)e(x)|p\rangle|mat\rangle\delta_{\alpha 0} = n_e(x)\delta_{\alpha 0}$ where $n_e(x)$ is the number density at the point x. Also, $\langle p|\overline{v_{eL}}(x)\gamma_\alpha v_{eL}|p\rangle = 1$

Substituting all these values, $H_I^{mat}(t) = \sqrt{2}G_F n_e(t)\beta$, Using $\beta_{v_e,v_e} = 1$, all other elements of matrix β are equal to zero.

Let us now consider the NC interaction. Induced by the $Z^0$ exchange, the Hamiltonian of NC interactions of neutrinos with electrons and nucleons has the form:

$$H_I^{CC}(x) = 2\frac{G_F}{\sqrt{2}} \sum_{l=e,\mu,\tau} \overline{v_{lL}}(x)\gamma^\alpha v_{lL}(x) j_\alpha^{NC}(x)$$

Where $j_\alpha^{NC}(x)$ is the sum of electron and nucleon (quark) neutral current. For the vector part of effective hadron neutral current,

$$v_\alpha^{NC(N)}(x) = \frac{1}{2}\overline{N}(x)\gamma_\alpha\tau_3 N(x) - 2\sin^2\theta_w \overline{p}(x)\gamma_\alpha p(x) \text{ As } N = \begin{pmatrix} p \\ n \end{pmatrix} \text{ and } \tau_3 = \begin{pmatrix} 1 & 0 \\ 0 & -1 \end{pmatrix}$$

where $\theta_w$ is the weak angle. Similarly for the effective part of electron current:

$$v_\alpha^{NC(n)}(x) = (-\frac{1}{2})\overline{n}(x)\gamma_\alpha n(x)$$

$$v_\alpha^{NC(e)}(x) = (-\frac{1}{2} + 2\sin^2\theta_w)\overline{e}(x)\gamma_\alpha e(x)$$

For the corresponding matter matrix elements we have

$$\langle mat|v_\alpha^{NC(e)}(x)|mat\rangle = (-\frac{1}{2} + 2\sin^2\theta_w)n_e(x)\delta_{\alpha 0}$$

$$\langle mat|v_\alpha^{NC(p)}(x)|mat\rangle = (\frac{1}{2} - 2\sin^2\theta_w)\rho_p(x)\delta_{\alpha 0}$$

and $\langle mat|v_\alpha^{NC(n)}(x)|mat\rangle = -\frac{1}{2}\rho_n(x)\delta_{\alpha 0}$

For the neutral matter, $n_e(x) = \rho_p(x)$, we conclude that the contributions of electron and proton NC to the effective Hamiltonian cancel each other. Thus

$$\langle mat|j_\alpha^{NC}(x)|mat\rangle = -\frac{1}{2}\rho_n(x)\delta_{\alpha 0}$$

By taking into account the effective charged current interaction, only $v_e$-e CC interaction gives a contribution to the effective Hamiltonian. Thus the evolution equation of neutrino has the form:

$$i\frac{\partial a(t)}{\partial t} = (U\frac{m^2}{2E}U^\dagger + \sqrt{2}G_F n_e(t)\beta)a(t)$$

Similarly for the antineutrinos effective Hamiltonian differs in sign from the neutrino-electron interactions, thus

$$\overline{H}_I^{mat}(x) = -\sqrt{2}G_F n_e(t)\beta$$

Here we are more concerned with matter effects of constant density. The total Hamiltonian of neutrino in matter

$H = U\frac{m^2}{2E}U^\dagger + \sqrt{2}G_F n_e\beta$ where $U = \begin{pmatrix} \cos\theta & \sin\theta \\ -\sin\theta & \cos\theta \end{pmatrix}$ then the total effective Hamiltonian is $H = \frac{1}{2}TrH + H^m$.

Here $\frac{1}{2}Tr\,H = \frac{m_1^2 + m_2^2}{4E} + \frac{1}{2}\sqrt{2}G_F n_e$ and $H^m$ is the traceless part of Hamiltonian.

$$H^m = \frac{1}{4E}\begin{pmatrix} -\Delta m^2 \cos 2\theta + A & \Delta m^2 \sin 2\theta \\ \Delta m^2 \sin 2\theta & \Delta m^2 \cos 2\theta - A \end{pmatrix} \text{ where } A = 2\sqrt{2}G_F n_e E$$

$$H^m = U^m E^m U^{m\dagger}$$

$$U^m = \begin{pmatrix} \cos\theta^m & \sin\theta^m \\ -\sin\theta^m & \cos\theta^m \end{pmatrix}$$

$$E^m = \begin{pmatrix} E_1^m & 0 \\ 0 & E_2^m \end{pmatrix}$$

Where $E_{1,2}^m = \pm\frac{1}{4E}\sqrt{(\Delta m^2 \cos 2\theta - A)^2 + (\Delta m^2 \sin 2\theta)^2}$

From equation, we find that the mixing angle $\theta^m$ is given as:

$$\cos 2\theta^m = \frac{\Delta m^2 \cos 2\theta - A}{\sqrt{(\Delta m^2 \cos 2\theta - A)^2 + (\Delta m^2 \sin 2\theta)^2}}$$

$$\sin 2\theta^m = \frac{\Delta m^2 \sin 2\theta}{\sqrt{(\Delta m^2 \cos 2\theta - A)^2 + (\Delta m^2 \sin 2\theta)^2}}$$

Three expressions for atmospheric neutrinos for normal hierarchy are given as follows:

$$P_{\mu\tau matter} = (Cos[\theta_{13}^m])^2 * (Sin[2\theta_{23}])^2 * (Sin[(1.27 * (\Delta m_{31} + A + \Delta m_{31}^m) * L)/E])\wedge 2 + (Sin[\theta_{13}{}^m])^2$$
$$* (Sin[2\theta_{23}])^2 * (Sin[(1.27 * (A + \Delta m_{31} - \Delta m_{31}^m) * L)/E])^2 - (Cos[2\theta_{23}{}^m])^2 * (Sin[2\theta_{13m}])^2$$
$$* (Sin[\theta_{23}])^2 * (Sin[(1.27 * \Delta m_{31}^m * L)/E])^2$$

$$P_{\mu\mu matter} = (1 - Cos[\theta_{13}^m])^2 * (Sin[2\theta_{23}])^2 * (Sin[(1.27 * (A + \Delta m_{31} - \Delta m_{31}^m) * L)/E])^2$$

$$P_{\mu e matter} = (Sin[2\theta_{13}{}^m])^2 * (Sin[\theta_{23}])^2 * (Sin[(1.27 * \Delta m_{31}^m) * L)/E])^2$$

And for inverted hierarchy

$$P_{\mu\tau matter} = (Cos[\theta_{13}^m])^2 * (Sin[2\theta_{23}])^2 * (Sin[(1.27 * (A - \Delta m_{31} + \Delta m_{31}^m) * L)/E])\wedge 2 + (Sin[\theta_{13}{}^m])^2$$
$$* (Sin[2\theta_{23}])^2 * (Sin[(1.27 * (A + \Delta m_{31} - \Delta m_{31}^m) * L)/E])^2 - (Cos[2\theta_{23}{}^m])^2 * (Sin[2\theta_{13m}])^2$$
$$* (Sin[\theta_{23}])^2 * (Sin[(1.27 * \Delta m_{31}^m * L)/E])^2$$

$$P_{\mu\mu matter} = (1 - Cos[\theta_{13}^m])^2 * (Sin[2\theta_{23}])^2 * (Sin[(1.27 * (A - \Delta m_{31} - \Delta m_{31}^m) * L)/E])^2$$

$$P_{\mu e matter} = (Sin[2\theta_{13}{}^m])^2 * (Sin[\theta_{23}])^2 * (Sin[(1.27 * \Delta m_{31}^m) * L)/E])^2$$

## 5. Oscillation parameters and masses:

Although atmospheric and solar neutrinos are having different origin yet they exhibit the same phenomenon of oscillations among their flavors. Till date, a vast range of experiments have been designed and performed to go deeper into details of various parameters involved in flavor transitions. Several experiments like SuperKamioKande [6] and SNO [7] have brought evidence for neutrino oscillations but the parameters involved in transitions still need to be more précised and accurate. The parameters involved in three flavor oscillations can be estimated from the neutrinos coming from the Sun, atmospheric and nuclear reactors and accelerators. Solar experiments detect neutrinos generated in the core of Sun due to thermonuclear reactions and must have the energy of the order of 0.2-15 MeV whereas atmospheric experiments detect neutrino produced in cascade initiated by Cosmic rays collisions with nuclei in the Earth's atmosphere and has a source detector at a distance of several tons of meter with a range L/E< 1m/MeV. Oscillation parameters historically fall into four categories in mixing angle-mass splitting parameter space: Vacuum oscillations (VAC), "LOW", small mixing angle (SMA), large mixing angle (LMA). Similarly based on distance from source to detector, detectors can be SBL, LBL, VSBL, VLBL. The era of neutrino oscillation experiments started with Chlorine –Homestake experiment [8] which used inverse Beta-decay to measure. Davis's Chlorine Home stake experiment was the first step and confirmed the discrepancy in number of Solar Neutrinos. Then a series of different experiments began to reach more close to the situation. Various experiments in this series include Kamiokande-II, Gallex and Sage, Super-Kamiokande, SNO, KAMLAND etc.. The KamioKande [9] measured solar neutrinos to be half as per the SM whereas two experiments Gallex and Sage [10] measured 56-60% of neutrino capture rate as predicted by standard model. Then Super-KamioKande I,II,III [11-13] provided evidence for non-zero mass and also produced observation consistent with μ-neutrinos changing into τ- neutrinos. SNO[14] detects $^8$B neutrinos via charged current interaction, neutral current interaction and elastic scattering. KAMLAND experiment was also able to investigate geographically produced anti-neutrinos [15] and the best fitted values of $\Delta m^2_{21}$= 7.58 + 0.14 (stat) $^{+0.15}_{-0.15}(syst) \times 10^{-5}$ eV$^2$ and $\tan \theta_{12} = 0.56 + ^{0.10}_{-0.07}(stat)\ ^{+0.10}_{-0.06}(syst)\ for\ \tan \theta_{12} < 1$ as calculated using KAMLAND experiment. K2K [16] was another long baseline experiment to study oscillation from $\nu_\mu$ to $\nu_e$ in the atmospheric region and confirmed the deficit of muon as observed in SuperKamioKande. MINOS [17] at Fermi National Laboratory studied muon oscillations produced from pion and kaon decay in the energy range of 1-10GeV and focused primarily on measurement of $\Delta m^2_{23}$ with the precision better than 10%.

A global analysis of different experiments can provide best fit values of different parameters. The two large mixing angles $\theta_{12}$ and $\theta_{23}$ have been found to be of the order of ~34$^0$ and ~ 45$^0$ respectively but the third mixing angle $\theta_{13}$ is suppressed by different experiments. Recent searches for $\theta_{13}$ however oppose suppression of this mixing angle. Moreover sign of $\Delta m^2_{31}$ is still unknown therefore two types of neutrino spectrum are possible, one is normal hierarchy (m$_1$<< m$_2$ < m$_3$) whereas other is inverted hierarchy which assume masses in order (m$_3$<< m$_1$ < m$_2$). Three mixing angles and the mass differences from global analysis of data [18] are given below:

$$\Delta m^2_{21} = 7.59 \pm 0.20\ ^{+0.61}_{-0.69} \times 10^{-5}\ eV^2$$

$$\Delta m^2_{31} = +2.46 \pm 0.12 (\pm 0.37) \times 10^{-3} eV^2$$

$$= -2.36 \pm 0.11 \pm 0.37 eV^2$$

$$\theta_{23} = 42.8^{+4.7}_{-2.9}(^{+10.7}_{-7.3})$$

$$\theta_{12} = 34.4 \pm 1.0 (^{+3.2}_{-2.9})$$
$$\theta_{13} = 5.6^{+3.0}_{-2.7}$$

### 6. Oscillation Plots for vacuum and matter:

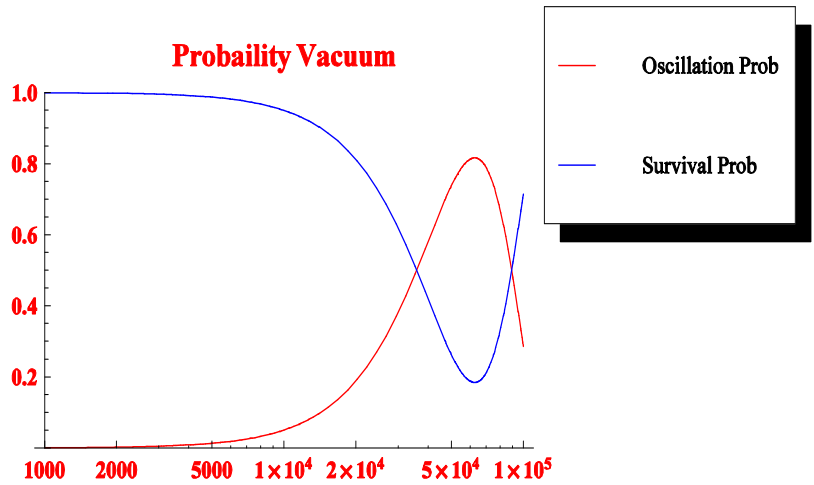

**Figure 5.1 Oscillations and Survival Prob**

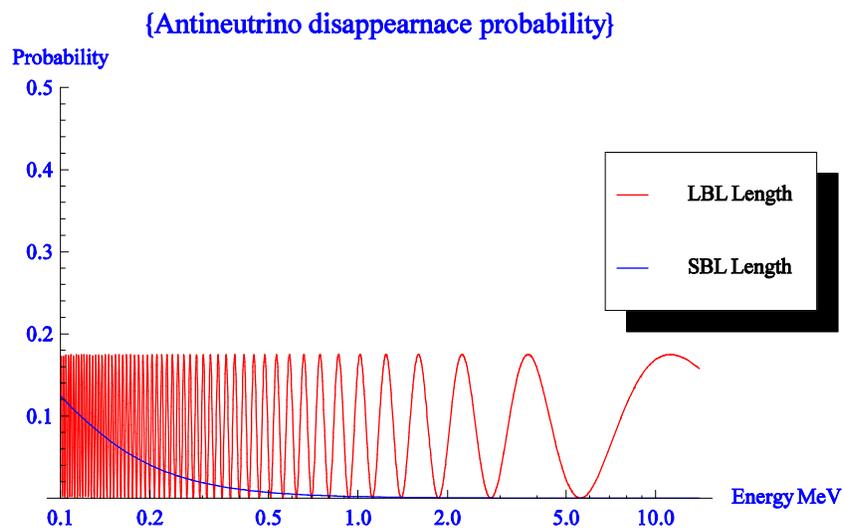

**Figure 5.2**

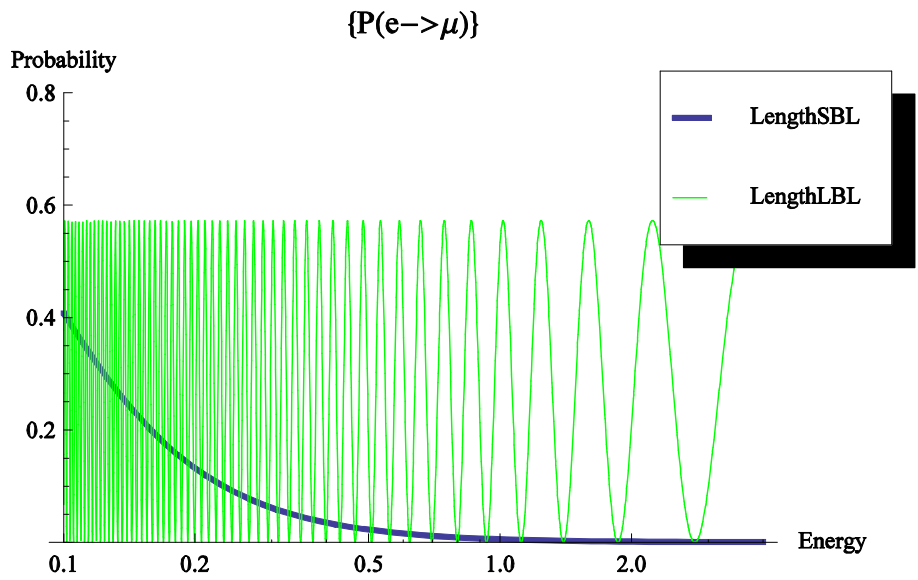

Figure 5.3

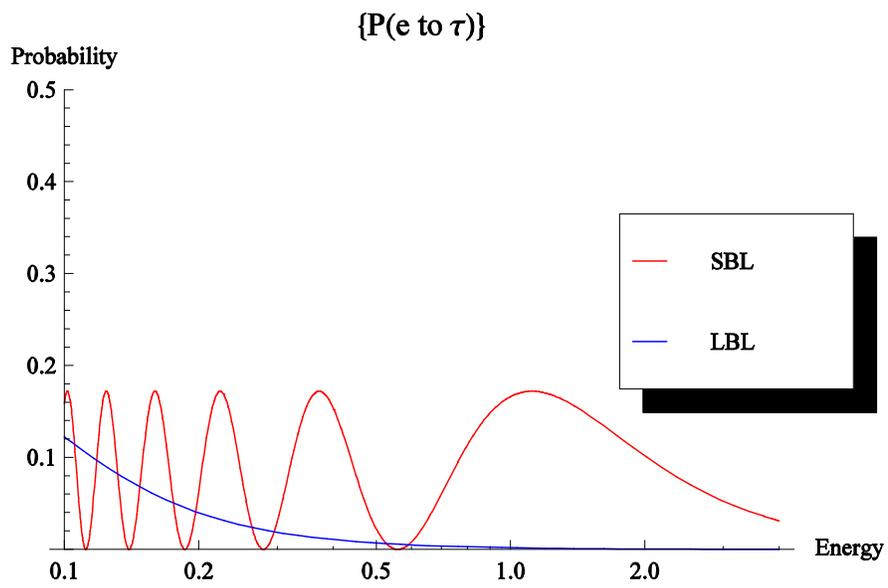

Figure 5.4

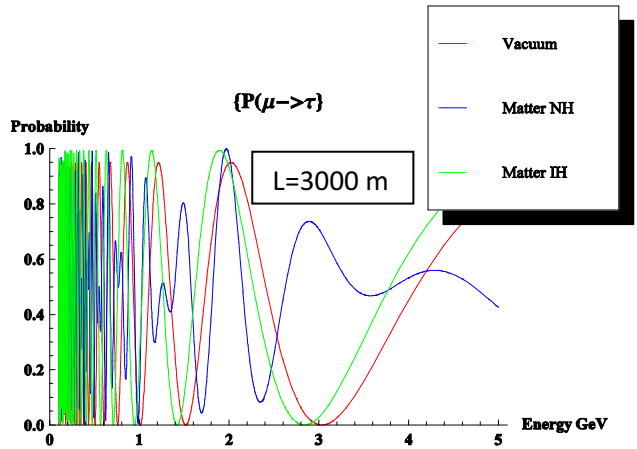

(a)

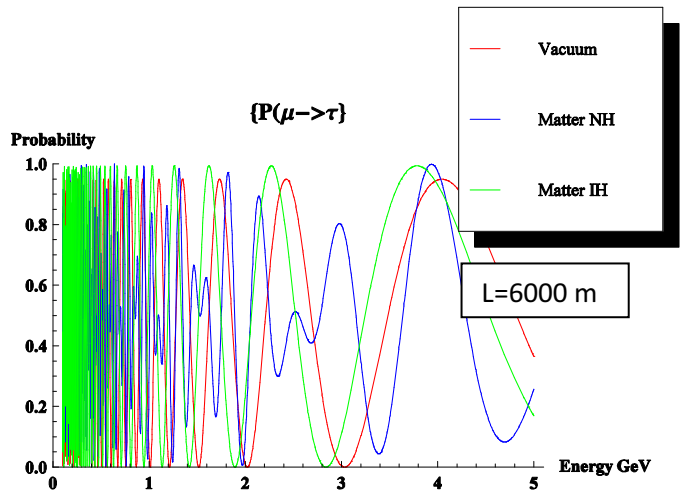

(b)

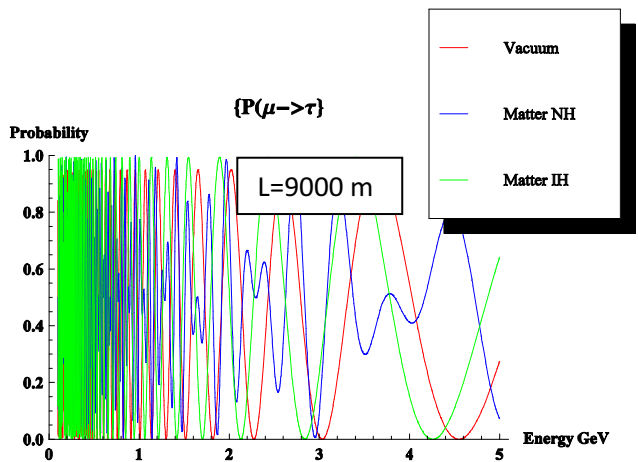

(c)

**Figure 5.5(a), (b), (c)**

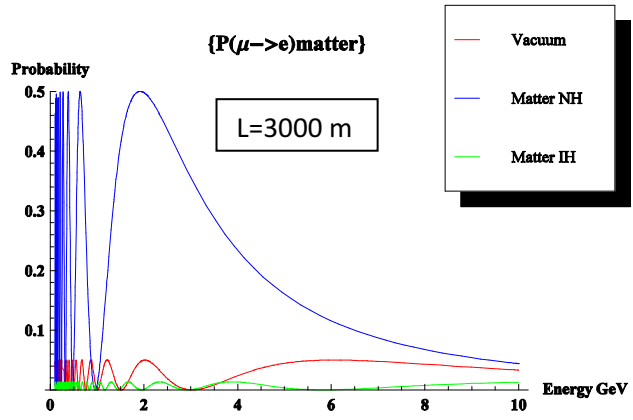

(a)

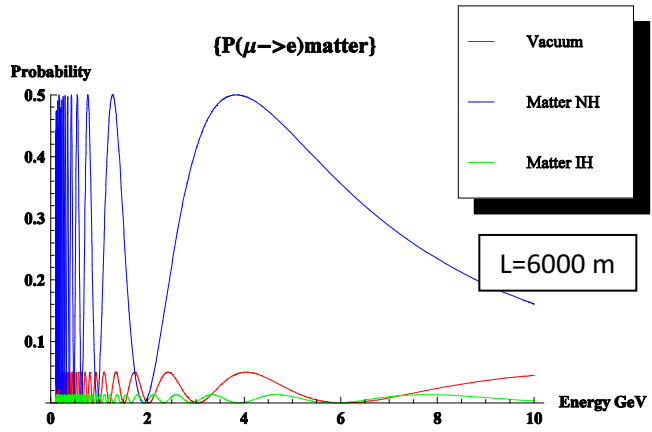

(b)

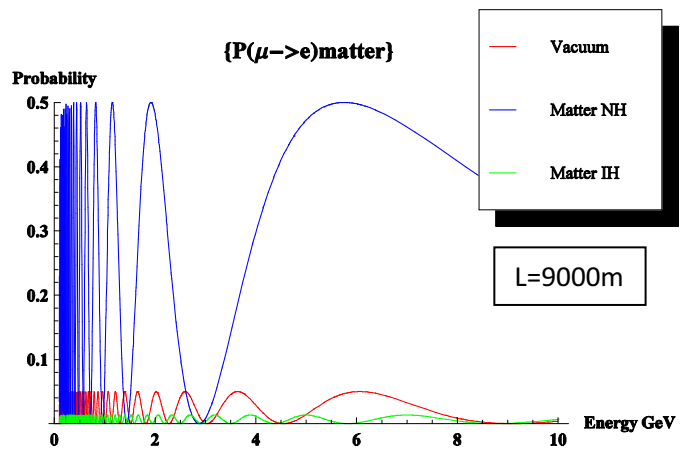

(c)

Figure 5.6(a),(b),(c)

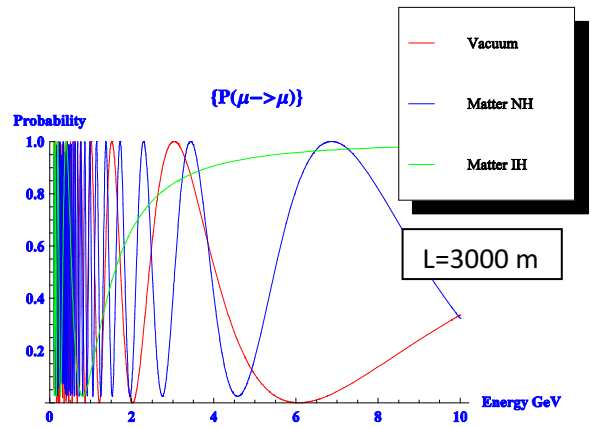

(a)

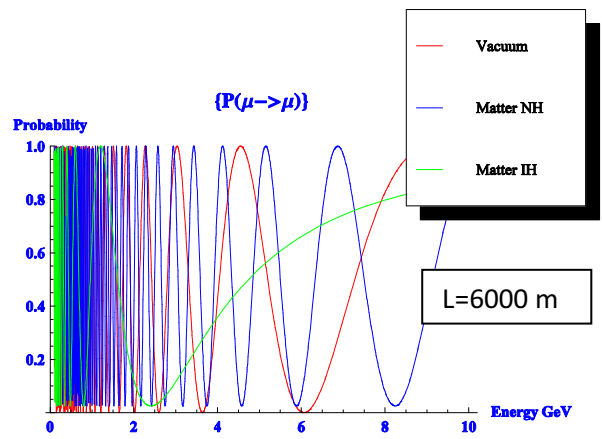

(b)

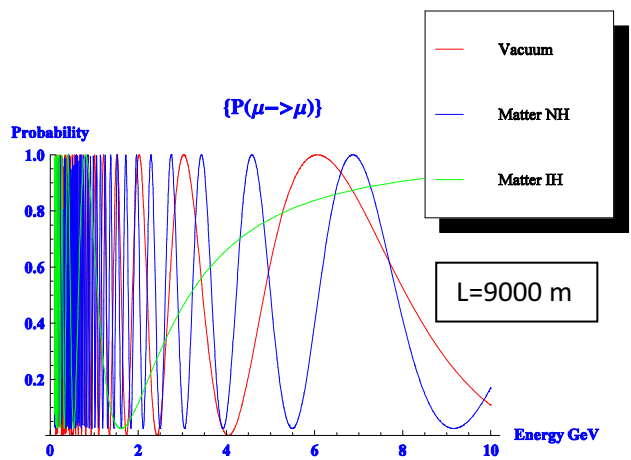

(c)

Figure 5.7(a),(b),(c)

## 7. Discussion of Plots and Conclusion:

The parameter affecting neutrino oscillation in matter is given by $A = 2\sqrt{2} G_F n_e E$ where $G_F$ is Fermi Constant, $n_e$ is the electron number density and E is energy of neutrino beam. The parameter A is almost constant for Long-baseline experiment with high precision as the beam does not penetrate deeply into earth without affecting the different layers of earth but in sun, density decreases exponentially from centre to core of earth which further leads to resonance enhancement of oscillations known as MSW effect [5]. We here restrict ourselves to matter effects with constant density for atmospheric neutrinos only. The graph for probability vs energy at three different lengths vacuum as well as in matter (normal and inverted hierarchy) for $\nu_\mu$ to $\nu_e$, $\nu_\mu$ to $\nu_\tau$, $\nu_\mu$ to $\nu_\mu$ shows that the presence of matter affects the neutrino oscillation probability the most in case of $\nu_\mu$ to $\nu_e$ oscillations. This may be due to reason that the charged current interaction with the electrons of the medium directly affects the oscillation probability in case of electrons only. This proves the fact that matter effects are flavor dependent. Moreover, oscillations become more frequent at low energy and increase in length results in increase of energy range which is more sensitive to oscillations but affects to a very lesser extent on magnitude of transition probability. Oscillation probability is more in magnitude in case of μ to τ rather than electron which concludes that a mass eigen state is more composed of μ and τ rather than electron for atmospheric neutrinos especially. The smallness of $\nu_e$ part of total mass eigen state is measured by the mixing angle $\theta_{13}$. For solar neutrino oscillation probability, reactor based neutrino experiments and detectors are designed as different experiments are designed in order to sensitive to different values of $\Delta m^2$, by choosing appropriate value of L/E, the neutrino beam consisting of $\nu_e$ when coming out of the source and having energy only ~1MeV that's why in order to satisfy $\Delta m^2 L/2E \sim 1$ and L can be of order such that L/E<= 1km/MeV. Solar neutrino experiments have sensitivity for $\Delta m^2$ of very small value. SNO has been extremely successful so far to explain and to prove neutrino flavor change as it was designed in such a way to detect the changed flavor of solar neutrinos that is the flavor in such a way to detect the changed flavor of solar neutrinos that is the flavor in which $\nu_e$ changed. After analyzing the results from flux ratios, it was found that day-night asymmetry has more sensitivity to value of $\Delta m^2_{12}$. The results favored large mixing angle solutions. CHOOZ [19] experiment put forward that electron antineutrino survival probability and its oscillation probability from electron to muon and tau are excluded for $\Delta m^2 \geq 8 \times 10^{-4}$ eV$^2$ and at maximum mixing $\sin^2 2\theta \geq 0.17$. The chain of experiment continued with KamLAND [20] experiment which provided evidence of antineutrino disappearance probability for the very first time with a long baseline length of the order of ~175 km as shown in figure 5.5. All reactor based neutrinos consider only vacuum part of the oscillation probability as matter effects are negligible at low energy. Transitions from electron to μ`s and τ`s are shown in figure 5.6 and 5.7. Two plots here shows that the maximum oscillations are from e to μ for solar neutrinos. Here the oscillation parameters have been taken from the particle data group [20].

Although fifty years of intense efforts have been made by researchers in various ways to understand this fundamental particle yet the information about this subatomic particle is not complete. Neutrino oscillations helped to solve the greatest mystery in solar world but actual mass of the neutrinos is still unknown. The recent searches for neutrinos are more focused on the estimation of parameter $\theta_{13}$ [21] and the speed of neutrinos. Some more pieces of information on the mass hierarchy and CP violating phase can help us to get more detail information on oscillation phenomenology.

In summary, two flavor oscillations are discussed here for probability measurements and moreover generalized to three flavor in vacuum and matter. In the era of neutrino physics, vast amount of the information for oscillation parameters we have in hand, still the future measurements are more sensitive to $\theta_{13}$. For solar neutrinos, a combined analysis of solar experiments can provide upper limits on $\theta_{13}$ upto a higher significance level. It is claimed that if $\theta_{13}$ is greater than the limit as expected then faster atmospheric oscillations can be predicted at about 1-2 km.